\documentclass[a4paper,11pt]{article}
\usepackage{pos}
\usepackage{cleveref}

\newcommand{\chpt}{\ifmmode{\mathrm{ChPT}}\else{ChPT}\fi}
\newcommand{\LO}{\ifmmode{\mathrm{LO}}\else{LO}\fi}
\newcommand{\NLO}{\ifmmode{\mathrm{NLO}}\else{NLO}\fi}
\newcommand{\NNLO}{\ifmmode{\mathrm{NNLO}}\else{NNLO}\fi}
\newcommand{\NNNLO}{\ifmmode{\mathrm{N^3LO}}\else{N$^3$LO}\fi}
\newcommand{\NkLO}[1][k]{\ifmmode{\mathrm{N}^{#1}\mathrm{LO}}\else{N\textsuperscript{\itshape #1}LO}\fi}

\newcommand{\Mpi}{M_\pi}
\newcommand{\Fpi}{F_\pi}
\newcommand{\Mpiz}{M_0}
\newcommand{\Fpiz}{F_0}

\DeclareMathOperator{\Tr}{Tr}
\newcommand{\p}{\partial}
\newcommand{\D}{\mathcal D}
\newcommand{\lagr}{\mathcal L}
\renewcommand{\O}{\mathcal O}
\newcommand{\const}{\text{const.}}

\newcommand{\diagram}[2][()]{\setbox0=\hbox{#1}\raisebox{-\dp0}{\includegraphics[height=\dimexpr\ht0+\dp0\relax]{diagrams/N3LO#2.pdf}}}

\usepackage{pgf,tikz}
\usetikzlibrary{shapes.geometric, decorations.text}

\tikzset{
    diagrams/.style 2 args={draw=#1!20, line width=3ex, ellipse, postaction={
          decorate, decoration=
            {text along path, raise=-.5ex, 
            text align=right, text color=#1,
            text={#2}}}}}

\title{The three-loop hadronic vacuum polarization in chiral perturbation theory}
\ShortTitle{The 3-loop HVP in ChPT}

\author*[a]{Mattias Sjö}
\author[a]{Laurent Lellouch}
\author[a]{Alessandro Lupo}
\author[b,c]{Kálmán Szabo}
\author[d]{Pierre Vanhove}

\affiliation[a]{Aix Marseille Univ, Université de Toulon, CNRS, CPT, Marseille, France}
\affiliation[b]{Department of Physics, University of Wuppertal, D-42119 Wuppertal, Germany}
\affiliation[c]{Jülich Supercomputing Centre, Forschungszentrum Jülich, D-52428 Jülich, Germany}
\affiliation[d]{Institut de Physique Théorique, Université Paris-Saclay, CEA, CNRS,\\F-91191 Gif-sur-Yvette Cedex, France}

\emailAdd{mattias.sjo@cpt.univ-mrs.fr}

\abstract{Hadronic vacuum polarization is a key observable in low-energy QCD, and is famously the greatest contributor to the theoretical uncertainty in the muon magnetic moment.
Its long-distance part in particular is a weak point of the current best lattice QCD computations.
In this summary of our recent work~\cite{Lellouch:2025rnz}, we present its computation to next-to-next-to-next-to-leading order in chiral perturbation theory, capturing the lowest-energy hadronic contributions to unprecedented precision and opening the door for improved control over lattice finite volume effects.
The result depends on a small number of low-energy constants, whose values are mostly under good control.
This calculation pushes the envelope of high-order chiral perturbation theory and of the evaluation of multiloop integrals with massive propagators, thereby extending the toolbox for precision calculations in very low-energy QCD.}

\FullConference{The 42nd International Symposium on Lattice Field Theory (LATTICE2025)\\
2-8 November 2025\\
Tata Institute of Fundamental Research, Mumbai, India\\}

\renewcommand{\hookAfterAbstract}{%
    \par\bigskip
    \textsc{ArXiv ePrint}: \href{https://arxiv.org/abs/2510.12885}{2510.12885}
}

\begin{document}
\maketitle

\begin{figure}[tp]
    \centering
    \includegraphics[width=.8\textwidth]{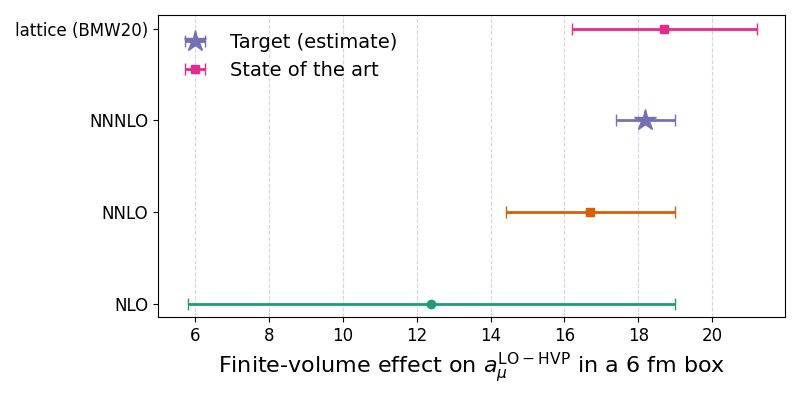}
    \caption{The FVE on the HVP contribution to the muon anomalous magnetic moment from \chpt\ and lattice QCD, including a projection of the ultimate result of this work.}
    \label{fig:guess}
\end{figure}

\section{Introduction}

Hadronic vacuum polarization (HVP) represents the low-energy QCD modifications to the propagation of photons in the vacuum, and its leading component can be intuitively understood as the appearance of virtual pion pairs.
This is understandably a much more suppressed effect than that of virtual electron-positron pairs, but despite being several orders of magnitude smaller than the QED contribution, HVP currently dominates the uncertainty of precision standard model observables such as the muon anomalous magnetic moment~\cite{Muong-2:2025xyk} or the value of the electromagnetic coupling at the electroweak scale~\cite{FCC:2018byv}, due to its nonperturbative nature making it much harder to compute.

There are two main approaches for computing HVP: data-driven, connecting it via dispersion relations to measured hadron production cross sections, and lattice QCD, where it is numerically simulated from first principles.
Recent years have seen the latter replace the former as the most trusted and precise standard model prediction~\cite{Aoyama:2020ynm,Aliberti:2025beg}, which puts pressure on the lattice side to further reduce the uncertainty.
A prominent source of systematic error is the finite volume of the lattice, which strongly affects the low-energy pion modes.
This has been given much attention in recent leading lattice calculations~\cite{Borsanyi:2020mff,Boccaletti:2024guq}.

HVP can also be computed using Chiral Perturbation Theory (\chpt), the effective field theory of low-energy QCD.
While \chpt\ lacks the predictiveness and physical scope for a directly competitive result, it excels in the energy range that dominates finite-volume effects (FVEs).
The difference between a \chpt\ observable in finite and infinite volume is therefore a good estimate of the FVE of the same observable on the lattice, and can be used to correct the associated error.
This has been done using \NLO\ (one-loop) \cite{Aubin:2015rzx} and \NNLO\ (two-loop) \chpt~\cite{Bijnens:2017esv,Aubin:2020scy,Borsanyi:2020mff}.
In ref.~\cite{Aubin:2020scy}, it was also proven that this approach is valid to arbitrarily high order, but doubt was expressed about whether higher-order results would ever be obtained.
In defiance of this, the present work presents the infinite-volume part of the \NNNLO\ (three-loop) calculation, laying the groundwork for the more challenging finite-volume one.
Assuming a geometric progression of errors, \NNNLO\ should be the order where a competitively precise FVE estimate is achieved; see \cref{fig:guess}.

\section{Theory}

\chpt\ is the low-energy effective theory obtained by breaking the chiral symmetry of massless QCD both spontaneously (via the quark condensate) and explicitly (via quark masses and non-QCD couplings, introduced as spurion fields).
We use the simple case of 2 isospin-symmetric flavors, a theory of a SU(2) triplet of pions of equal mass, which we couple to a non-dynamic photon field.
The Lagrangian is arranged following a formal power-counting scheme, with the LO part being
\begin{equation}
    \lagr_\LO = \frac{\Fpiz^2}{4} \Tr\big[D_\mu U (D^\mu U)^\dag + \Mpiz^2(U^\dag + U)\big]\,,
\end{equation}
where $\Mpiz,\Fpiz$ are the bare pion mass and decay constant, $D_\mu = \p_\mu - iA_\mu[Q, U]$ is the covariant derivative introducing the photon $A_\mu$ via the quark charge matrix $Q$, and $U$ is a unitary matrix parametrizing the pions $\{\phi^1,\phi^2,\phi^3\}$ as
\begin{equation}\label{eq:param}
    U = \frac{i\Phi}{\Fpiz\sqrt2} + \O(\Phi^2)\,,\qquad
    \Phi = \sum_{i=1}^3 \sigma^i \phi^i\,,
\end{equation}
where the algebra is normalized with $\Tr(\sigma^i\sigma^j) = \delta^{ij}$.
This is followed by a tower of higher-order Lagrangians, with $\lagr_{\NkLO}$ containing the counterterms needed for renormalization starting at $k$ loops.
In our case, there are 7 counterterms at \NLO~\cite{Gasser:1983yg}, 57 at \NNLO~\cite{Bijnens:1999hw,Bijnens:1999sh}, and 475 at \NNNLO~\cite{Bijnens:2018lez}, each associated with an \emph{a priori} unknown low-energy constant (LEC).
The bare mass and decay constant relate to their physical counterparts $\Mpi,\Fpi$ via
\begin{equation}\label{eq:massdecay}
    \Mpiz^2 \to \Mpi^2\bigg[1 + \sum_{i=1}^\infty \xi^i \delta M_i\bigg]\,,\qquad
    \Fpiz   \to \Fpi  \bigg[1 + \sum_{i=1}^\infty \xi^i \delta F_i\bigg]\,,\\
\end{equation}
where $\xi=\frac{\Mpi^2}{16\pi\Fpi^2}$ and the coefficients $\delta M_i,\delta F_i$ can be found in ref.~\cite{Bijnens:2017wba}.

\begin{figure}
    \begin{tikzpicture}
        \begin{scope}[xshift=.5cm]
            \node[diagrams={blue}{NLO contact~~~~}]
                (N10) at (0,0) {\includegraphics[scale=1.5]{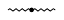}};
            \node[right of=N10, anchor=west, outer sep=1cm, inner sep=0cm,
                diagrams={green!70!black}{\NLO\ 1-loop~~}]
                (N11) {\includegraphics[scale=1.5]{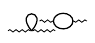}};
        \end{scope}
        \begin{scope}[xshift=-2cm, yshift=-2.5cm]
            \node[diagrams={blue}{NNLO contact~~~}]
                (N20) at (0,0) {\includegraphics[scale=1.5]{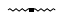}};
            \node[right of=N20, anchor=west, outer sep=1cm, inner sep=0cm,
                diagrams={green!70!black}{NNLO 1-loop}]
                (N21) {\includegraphics[scale=1.5]{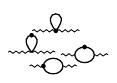}};
            \node[right of=N21, anchor=west, outer sep=2cm, inner sep=-.3cm,
                diagrams={yellow!70!black}{NNLO 2-loop}]
                (N22) {\includegraphics[scale=1.5]{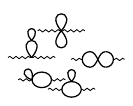}};
        \end{scope}
        \begin{scope}[yshift=-6cm]
            \node[diagrams={blue}{NNNLO contact~~}]
                (N30) at (0,0) {\includegraphics[scale=1.5]{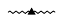}};
            \node[right of=N30, anchor=west, outer sep=1cm, inner sep=-.4cm,
                diagrams={green!70!black}{NNNLO 1-loop}]
                (N31) {\includegraphics[scale=1.5]{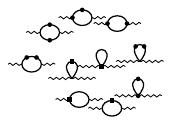}};
            \node[below of=N30, anchor=north, outer sep=.2cm, inner sep=-.7cm,
                diagrams={yellow!70!black}{NNNLO 2-loop}]
                (N32) {\includegraphics[scale=1.5]{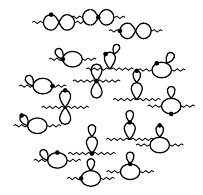}};
            \node[right of=N32, anchor=west, outer sep=2.2cm, yshift=-1cm, inner sep=-.7cm,
                diagrams={orange}{NNNLO 3-loop, factorizable}]
                (N33) {\includegraphics[scale=1.5]{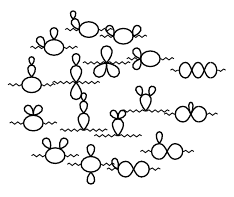}};
            \node[above right of=N33, anchor=south west, outer sep=3cm, xshift=-1cm,inner sep=-.2cm,
                diagrams={red}{NNNLO 3-loop, elliptic}]
                (N3E) {\includegraphics[scale=1.5]{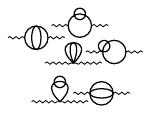}};
        \end{scope}
    \end{tikzpicture}
    \caption{
        All diagrams used in the HVP calculation, categorized by power-counting order and number of loops.
        Dots, squares and triangles represent \NLO, \NNLO\ and \NNNLO\ counterterms, respectively.}
    \label{fig:diagrams}
\end{figure}

Due to $G$-parity, all \chpt\ vertices connect an even number of pion lines.
There may additionally be one or two photon lines, with the simplest vertices being indistinguishable from those of scalar QED.
Stating the power-counting scheme informally, an \NkLO\ amplitude is built from Feynman diagrams with up to $k$ loops, with $k'$ loops being exchangeable for a counterterm from the \NkLO[$k'$] Lagranginan, or for mass and decay constant renormalization up to $\O(\xi^{k'})$.
Following this, the HVP diagrams are shown in \cref{fig:diagrams}.
These diagrams were computed using the new \texttt{FORM}~\cite{Vermaseren:2000nd,Ruijl:2017dtg} library \texttt{ChPTlib}~\cite{chptlib}.
This and other implementation code can be found in the attached repository~\cite{hvpfiles}.

\section{Loop integrals}

The main challenge in this calculation is its loop integrals.
As indicated in \cref{fig:diagrams}, most 3-loop diagrams (and all 1- and 2-loop ones) are of the factorizable kind, i.e., their momenta can be routed so that no propagator carries more than one loop momentum, and their loop integrals can consequently be decomposed as products of easily solvable 1-loop tadpole and bubble loops.
However, the six 3-loop diagrams highlighted in red are not factorizable, and in fact correspond to a particularly difficult class of loop integrals that cannot be expressed in terms of logarithms, polylogarithms, or even multiple polylogarithms, but which require the use of elliptic functions.
We will now go through the set of techniques we use to tackle these integrals.

The basic tool is integration-by-parts (IBP) reduction.
That is, if $\{\ell_i\}$ are the loop momenta, $I_n$ some dimensionally regularized integral, and $J_n$ its integrand---some product of powers of propagators---then IBP states that, for any momentum $q^\mu$ and index $j$,
\begin{equation}
    I_n = \bigg[\prod_i\int\frac{\mathrm d^d \ell_i}{\pi^{d/2}}\bigg] J_n
    \quad\Rightarrow\quad
    0 = \bigg[\prod_i\int\frac{\mathrm d^d \ell_i}{\pi^{d/2}}\bigg] \frac{\p (q^\mu J_n)}{\p\ell_j^\mu}\,.
\end{equation}
This modified integrand $\p (q^\mu J_n)/\p\ell_j^\mu$ is expressible as a linear combination of products of powers of propagators, so one finds an equation of the form $0 = \sum_m \alpha_m I_m$, where $\{I_m\}$ are different integrals and $\alpha_m$ are functions of the external kinematics and the dimension $d$.
By varying $n,j$ and $q^\mu$, this gives a system of equations that may be systematically solved, allowing any set of integrals to be reduced to a finite~\cite{Smirnov:2010hn} irreducible set known as the \emph{master integrals}.
This reduction is done via the Laporta algorithm~\cite{Laporta:2000dsw}, of which we use the improved implementation in the Mathematica package \texttt{LiteRed 2}~\cite{Lee:2012cn,Lee:2013mka}.
In our case, we have six elliptic master integrals, which we label as $E_{1,\ldots,6}(d;t)$ where $t$ is the photon virtuality.
$E_1$ is the basic integral associated with the graph \diagram{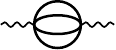}, $E_4$ the one for \diagram{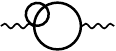}, and $E_5$ the one for \diagram{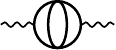}; $E_2$ ($E_3$) are obtained by squaring (cubing) a propagator in $E_1$, and $E_6$ by squaring one in $E_5$.

Our master integrals are divergent in four dimensions, but there exists a trick due to Tarasov~\cite{Tarasov:1996br} that allow them to be replaced by their finite two-dimensional counterparts.
It takes the form of an easily obtained differential operator $\D$ such that
\begin{equation}
    \D E_n(d;t) = E_n(d-2;t)\,,
\end{equation}
where the left-hand side can be IBP-reduced to a linear combination of $d$-dimensional master integrals.
Inverting this relation allows any four-dimensional master integral to be expressed as a linear combination of two-dimensional ones, thus separating the concern of divergence from that of integration.
Serendipitously, it also eliminates one of the master integrals: $E_4(4;d)$ is a linear combination of $E_{1,2,3}(2;d)$ and factorizable integrals, but does not contain $E_2(2;d)$.

A natural hierarchy exists among the master integrals based on which propagators they feature.
In our case, it looks like
\begin{equation}
    \{E_5,E_6\} > \{E_4\} > \{E_1, E_2, E_3\} > \{(\text{non-elliptic})\}\,,
\end{equation}
where $A > B$ means that the set of propagators in $B$ is a subset of those of $A$, and we say that $B$ is lower in the hierarchy.
Integrals on the same level have different powers of the same propagators.
A consequence of this is that the derivative of a master integral is a linear combination of master integrals at the same or lower level in the hierarchy, and a consequence of that is that each master integral satisfies a $n$th-order inhomogeneous differential equation, where $n$ is the number of integrals on its level in the hierarchy, and the inhomogeneous part contains integrals lower in the hierarchy.
Remarkably, in the case of $E_1$ its equation
\begin{equation}
    \bigg[t^2(t-4)(t-16)\frac{d^3}{dt^3} + 6t(t^2-15t+32)\frac{d^2}{dt^2} + (7t^2-68t+64)\frac{d}{dt} + (t-4)\bigg] E_1(2;t) = -12
\end{equation}
has a known solution in terms of elliptic polylogarithms~\cite{Bloch:2014qca}.
From this follows solutions for $E_{2,3}(2;t)$ via
\begin{equation}
    E_2(2;t) = -\frac14\bigg[t\frac{d}{dt} + 1\bigg] E_1(2;t)\,,\qquad
    E_3(2;t) = \bigg[\frac{t}{2}\frac{d^2}{d t^2} + \frac{4+t}{8}\frac{d}{d t}  + \frac{1}{8} \bigg] E_1(2;t)\,,
\end{equation}
which via Tarasov's relation gives $E_{1,\ldots,4}(4;t)$.
Some additional infrastructure is needed to express these elliptic objects in terms of real kinematics, which is covered in our upcoming work~\cite{LLSV}.

The last two master integrals are more problematic, since the coefficients that relate $E_{5,6}(d;t)$ to $E_n(d-2;t)$ diverge as $d\to4$.
Therefore, while for $n=1,\ldots,4$ we schematically have
\begin{equation}
    E_n(4-2\epsilon;t) = \frac{\const}{\epsilon^3} + \frac{[F(t)]}{\epsilon^2} + \frac{[F(t)]}{\epsilon} + [F, E_{1,2,3}(2;t)] + \O(\epsilon)
\end{equation}
where $[\ldots]$ stands for ``some linear combination of $\ldots$'' and $F(t)$ stands for ``all finite non-elliptic integrals'', for $n=5,6$ we have
\begin{equation}
    E_n(4-2\epsilon;t) = \frac{\const}{\epsilon^3} + \frac{[F(t), E_{1,2,3}(2;t)]}{\epsilon^2} + \frac{[F(t), E_{1,2,3}(2;t)]}{\epsilon} + [F, E_{1,\ldots,6}(2;t)]  + \O(\epsilon)\,.
\end{equation}
This breaks renormalization, since the sum of all counterterms is necessarily
\begin{equation}
    \frac{\const}{\epsilon^3} + \frac{[F(t)]}{\epsilon^2} + \frac{[F(t)]}{\epsilon} + \O(1)\,,
\end{equation}
and no linear combination of non-elliptic functions can cancel an elliptic function, as these are fundamentally different mathematical objects.
Thus, renormalizability demands the existence of some hidden relation that makes the elliptic integrals cancel among themselves.
This is unexpected, since there are normally no relations beyond those given by the IBP reduction.

Non-IBP relations have been previously encountered in sunset integrals with independent masses~\cite{Remiddi:2013joa}, and can be derived from the Gram matrix $G_{ij} = \ell_i\cdot\ell_j$.
Due to linear dependence, $|G|=0$ in $d=2$, which is known as the Schouten identity.
Thus, $G$ inserted as numerator in an integral will, after IBP reduction, give a linear combination of master integrals that vanishes in $d=2$, and if correctly constructed, this vanishing will not be manifest under $d$-dimensional IBP.
This is the simplest instance of what we call \emph{Schouten relations}, and the ones needed to restore renormalization in our case are more complicated, involving not just $E_n(2;t)$ but also $\frac{d}{d\epsilon} E_n(2-2\epsilon;t)$.
As a consequence, it becomes easier to work with $\bar E_{5,6}(4;t)$, which is the finite part of $\bar E_{5,6}(4-2\epsilon;t)$ as $\epsilon\to0$, rather than  $E_{5,6}(2;t)$ which become rather mangled by the relations.
We can still use differential equations, namely
\begin{align}
    (t-4)\bigg[\frac{(t-4)t^2}{4}\frac{d^2}{d t^2} + \frac{(t-3)t}{2}\frac{d}{d t} + \frac{1}{2}\bigg]\bar E_5(4;t) &= [F(t), E_{1,2,3}(2;t)]\,,\\
    \bar E_6(4;t) + \bigg[\frac{d}{dt}+\frac1t\bigg]\bar E_5(4;t) &= [F(t), E_{1,2,3}(2;t)]\,.
\end{align}
Given our closed-form expressions for $E_{1,2,3}(2;t)$, these equations lend themselves to a high-precision numerical solution, which we describe in forthcoming work~\cite{LLSV}.

\section{Results}

We denote by $\Pi^{\mu\nu}(t)$ the hadronic part of the photon two-point function, i.e., HVP, where $t=q^2/\Mpi^2$ is the normalized virtuality associated with the photon momentum $q^\mu$.
Due to Lorentz invariance, it can be decomposed as
\begin{equation}
    \Pi^{\mu\nu}(q) = (q^\mu q^\nu - q^2\eta^{\mu\nu}) \Pi_T(q^2) + q^\mu q^\nu \Pi_L(t)\,,
\end{equation}
and a powerful sanity check of our calculations is that $\Pi_L(t)=0$ as required by the Ward--Takahashi identity.
Other checks include the success of renormalization and invariance under changing the $\O(\Phi^2)$ part of \cref{eq:param}, which is a nontrivially manifesting symmetry of \chpt; together, these essentially rule out the possibility of calculation mistakes.
We decompose the on-shell subtracted transverse part as
\begin{equation}
    16\pi^2\big[\Pi_T(t) - \Pi_T(0)\big] = \bar\Pi_T^\NLO(t) + \xi\bar\Pi_T^\NNLO(t) + \xi^2\bar\Pi_T^\NNNLO(t) + \O(\xi^3)\,,
\end{equation}
where $\xi\approx0.014$ is the effective power-counting parameter introduced in \cref{eq:massdecay}.
Then the \NLO~\cite{Gasser:1983yg} and \NNLO~\cite{Golowich:1995kd} HVP is
\begin{equation}
    \bar\Pi_T^\NLO(t) = 2B(t) + \tfrac23\,,\qquad
    \bar\Pi_T^\NNLO(t) = tB(t)^2 - 4tB(t) l_6 - 8c_{56} t\,,
\end{equation}
where $l_i$ and $c_i$ are the renormalized \NLO\ and \NNLO\ LECs, respectively, and
\begin{equation}
    B(t) = -\frac19 + \frac{4-t}{3t}\int_0^1 \log\big[1 - x(1-x)t\big] \mathrm d x
\end{equation}
expresses the basic one-loop integral.
For brevity, we have simplified the renormalization scheme to omit logarithms of the renormalization scale, so the LECs as they appear here are not directly compatible with the literature.
In this same abbreviated form, our main result is
\begin{multline}\label{eq:main}
    \bar\Pi_T^\NNNLO(t) = E(t)
        + tB(t)^2\big[t(l_2 - 2l_1 - 2l_6) + 2l_4\big]
        + 2tB(t) \big[(4l_4 - tl_6)l_6  + r_{V1} + t r_{V2}\big]
        \\- 16t l_4 c_{56}
        + 16t(\tilde c_{332} - \tilde c_{333}) - 8t^2 \tilde c_{459}\,,
\end{multline}
where $\tilde c_i$ are \NNNLO\ LECs, $r_{V1} = -4(2c_{53}+c_{35}+4c_{6})$, $r_{V2} = (c_{53}-c_{51})$, and $E(t)$ is a lengthy linear combination of $E_{1,2,3}(2;t)$, $\bar E_{5,6}(4;t)$, $B(t)$ and higher polylogarithms.
The values of $r_{V2}$ and all \NLO\ LECs are known~\cite{Bijnens:2014lea}, but the others are not.
However, terms not containing $B(t)$ or $E(t)$ [in particular, the second line of \cref{eq:main}] have limited physical impact, contributing neither to cuts nor to FVEs, so only $r_{V1}$ need be determined for these.
$r_{Vi}$ are connected to the pion charge radius, which is the subject of ongoing studies; see, e.g., ref.~\cite{Colangelo:2021moe}.

\section{Conclusions}

This work represents the second-ever three-loop \chpt\ amplitude (the first being ref.~\cite{Bijnens:2017wba}) and a major step toward the extension of the HVP FVE program~\cite{Aubin:2015rzx,Bijnens:2017esv,Aubin:2020scy,Borsanyi:2020mff} to \NNNLO.
With the upcoming publication of ref.~\cite{LLSV}, this amplitude is fully under control in the continuum, and can serve as a blueprint for its finite-volume counterpart.
Reaching this point required a certain amount of pushing the envelope in terms of loop integral technology, something that is always useful for perturbation theory as a whole, since the same integrals may show up in completely different phenomenological settings---for instance, those needed for~\cite{Bijnens:2017wba} were already known from gauge theory~\cite{Melnikov:2000zc}, despite the fact that the masslessness of gauge bosons usually gives rather different master integrals than what one encounters in \chpt.
Lastly, we note that, while the predictivity of \chpt\ can be quite limited at this high order, our result is expressible in terms of rather few LECs, most of which are known; for the unknown ones, the multitude of phenomenological pathways to which this amplitude connects might provide a means for their determination.

\begin{figure}
    \centering
    \includegraphics[width=.3\textwidth]{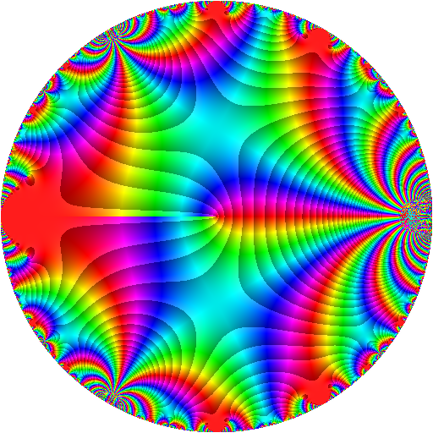}
    \quad
    \includegraphics[width=.3\textwidth]{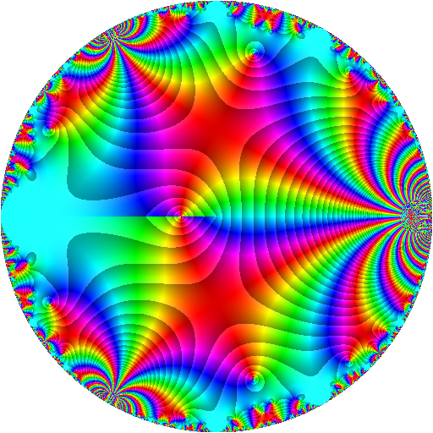}
    \quad
    \includegraphics[width=.3\textwidth]{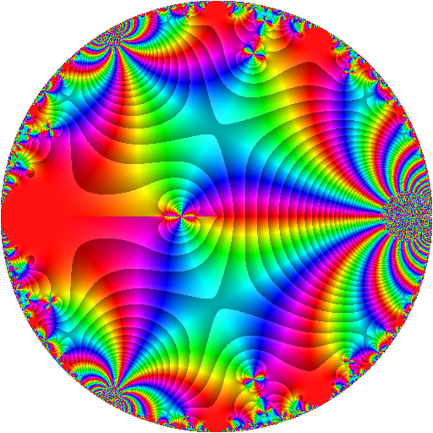}
    \caption{$E_{1,2,3}(2;t)$, domain-colored in their native elliptic habitat.
        Each point in the complex $t$-plane is represented infinitely many times in each circle, but you will find one instance of $t=\infty$ at the center of the circle, $t=16$ (the four-pion threshold) just to the left of that where $E_2(2;t)$ goes singular, $t=4$ (the two-pion threshold) at the large saddle point above, and $t=0$ on the rightmost edge, amidst the raging essential singularity.}
    \label{fig:masters}
\end{figure}

As a closing reflection on the singularly beautiful setting of LATTICE2025, we wish to conclude with a floral arrangement of our master integrals in \cref{fig:masters}.

\subsection*{Acknowledgments}

We thank Maarten Golterman and Martin Hoferichter for informative discussions after the talk on which these proceedings are based.
This work was supported by the French government under the France 2030 investment plan as part of the ``Initiative d'Excellence d'Aix-Marseille Université -- A\kern-2pt*MIDEX'' under grant AMX-22-RE-AB-052, by the Agence Nationale de la Recherche (ANR) under the grants ANR-22-CE31-0011 and ANR-24-CE31-7996, and by the Munich Institute for Astro-, Particle and BioPhysics (MIAPbP) which is funded by the Deutsche Forschungsgemeinschaft (DFG, German Research Foundation) under Germany's Excellence Strategy -- EXC-2094 -- 390783311.

\bibliographystyle{JHEP}
\bibliography{references}

\end{document}